\documentclass[a4paper,11pt]{article}
\usepackage{pos}

\usepackage{graphicx}
\usepackage{subcaption}
\usepackage{lipsum}
\usepackage{amsmath}
\usepackage[normalem]{ulem}
\usepackage{bigints}
\usepackage{slashed}

\newcommand{\cf}[1]{{Fig.~\ref{#1}}}

\newcommand{\msbar} {\overline{\text{MS}}}
\usepackage{feynmp}
\usepackage{mathpazo}
\usepackage{wrapfig}

\title{Revisiting inclusive production of $J/\psi$ and $\Upsilon$ in high-energy $\gamma\gamma$ collisions}

\author*[a,b]{Yelyzaveta Yedelkina}
\author[a]{Jean-Philippe Lansberg}
\author[a]{Maxim Nefedov}

\affiliation[a]{IJCLab, Université Paris-Saclay,\\
15, rue Georges Clémenceau, 91406, ORSAY Cedex, FRANCE}

\affiliation[b]{Physics School, University College Dublin (UCD),\\
            Science Buildings, Belfield, Dublin 4, Ireland}

\emailAdd{yelyzaveta.yedelkina@ijclab.in2p3.fr}
\emailAdd{Jean-Philippe.Lansberg@ijclab.in2p3.fr}
\emailAdd{maxim.nefedov@ijclab.in2p3.fr}

\let\OLDthebibliography\thebibliography
\renewcommand\thebibliography[1]{
  \OLDthebibliography{#1}
  \setlength{\parskip}{0pt}
  \setlength{\itemsep}{-1.pt}
}
\abstract{The impact of Next-to-Leading Order (NLO) QCD corrections to the differential distributions of $J/\psi$ and $\Upsilon$ mesons produced inclusively in $\gamma\gamma$ collisions is discussed for the kinematical conditions of LEP II for DELPHI and at the future Circular Electron-Positron Collider (CEPC). We take into account all sizeable contributions at LO in $v^2$ in NRQCD factorisation: 1) pure QED process $\gamma + \gamma \rightarrow J/\psi({}^3S^{[1]}_1) + \gamma$ up to $\alpha^3\alpha_s$, 2) single-resolved-photon contributions up to $\alpha \alpha_s^3$, 3) $\gamma+\gamma\to J/\psi + c\bar{c}$ up to $\alpha^2\alpha_s$ and 4) $\gamma+\gamma\to J/\psi + ggg$ up to $\alpha^2\alpha_s^3$. We will also discuss the pure QED process as a contribution to the exclusive production in ultra-peripheral heavy-ion collisions (UPC) at the LHC.}

\FullConference{
  The European Physical Society Conference on High Energy Physics (EPS-HEP),\\
  20-25 August 2023\\
  DESY and Universität Hamburg, Hamburg, Germany
}

\begin{document}
\maketitle

\vspace*{-0.6cm}
\section{Introduction: inclusive quarkonium production in $\gamma \gamma$ fusion}
\vspace*{-0.1cm}

Heavy quarkonia ($\cal{Q}$) are bound states of heavy quark-antiquark pair, $Q\overline{Q}$ with $Q=c,b$. The study of quarkonium production at lepton-lepton, hadron-hadron and lepton-hadron colliders allows us to extend our knowledge of quantum chromodynamics (QCD). Particularly, it allows one to learn more on the interplay between perturbative and non-pertubative QCD, namely the mechanism of perturbative creation of the $Q\overline{Q}$ pair and its non-perturbative hadronisation of $Q\overline{Q}$ into $\cal{Q}$. Here we discuss the production of vector quarkonium states $J/\psi$ and $\Upsilon$ -- spin-1 bound states of $c\bar{c}$ and $b\bar{b}$ quarks.

\vspace*{-0.45cm}
\begin{figure}[h!]
    \centering
 \begin{subfigure}[b]{0.17\textwidth}
  \centering
  \includegraphics[width=1\textwidth]{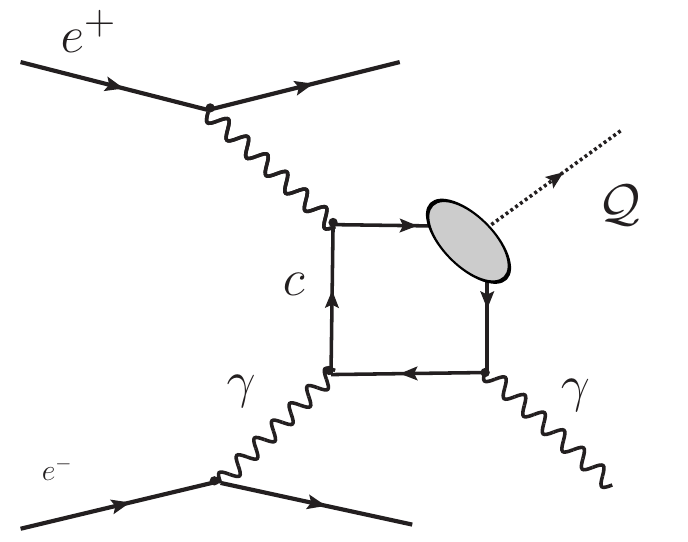}\vspace*{-0.25cm}
  \caption{   }
  \label{direct}
\end{subfigure}
\begin{subfigure}[b]{0.17\textwidth}
  \centering
  \includegraphics[width=1\textwidth]{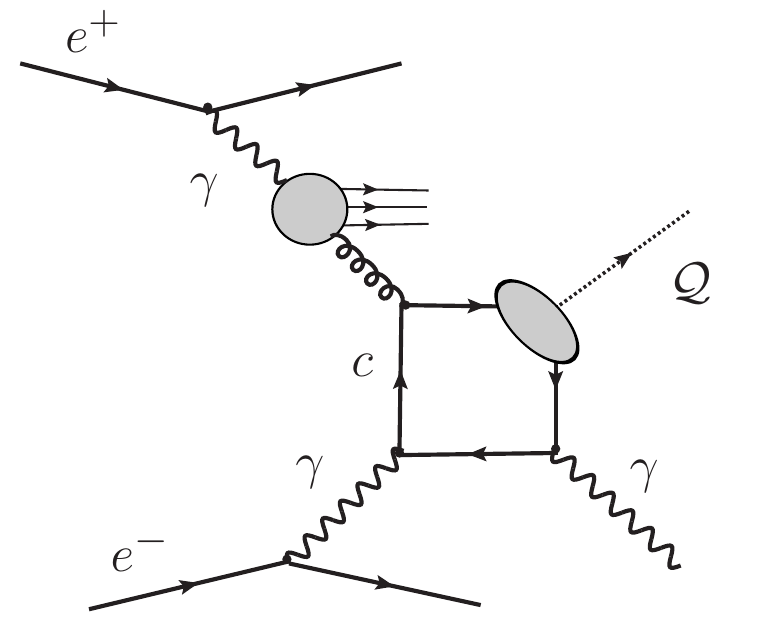}\vspace*{-0.25cm}
  \caption{   }
  \label{resolved}
\end{subfigure}
\begin{subfigure}[b]{0.17\textwidth}
  \centering
  \includegraphics[width=1\textwidth]{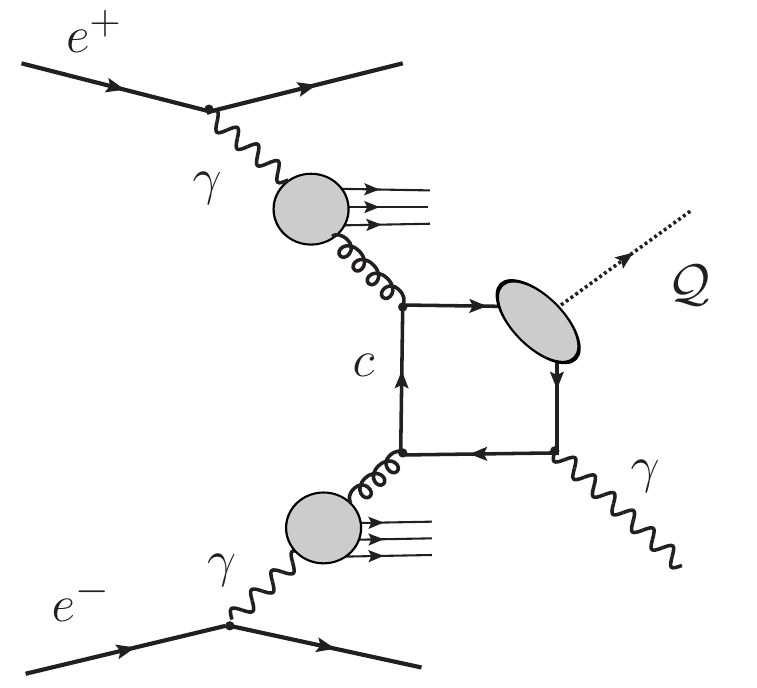}\vspace*{-0.25cm}
  \caption{   }
  \label{dresolved}
\end{subfigure}\vspace*{-0.2cm}
\caption{An illustrative selection of Feynman diagrams for inclusive $\cal{Q}$ production in $\gamma\gamma$ fusion 
via direct-photon (\ref{direct}), single-resolved-photon (\ref{resolved}) and double-resolved-photon (\ref{dresolved}) processes. 
}
\end{figure}
\vspace*{-0.15cm}

In this work, we discuss two classes of mechanisms which contribute to the inclusive $J/\psi$($\Upsilon$) production in $e^+e^-$ annihilation: direct-photon (\cf{direct}) and single-resolved-photon (\cf{resolved}) production in $\gamma \gamma$ fusion. In direct-photon production, the electron and the positron each emits a quasi on-shell photon which interact to produce the quarkonium. In single-resolved-photon $J/\psi$($\Upsilon$) production, one of the emitted quasi on-shell photon interacts with the second photon via its quark and gluon content. Double-resolved-photon production (\cf{dresolved}) was found to be at the per cent level by \cite{Klasen:2001cu} and we will not consider it here.

Nowadays, there is no consensus in the theory community on which mechanism is dominant in the hadronisation of the quarkonium produced in a hard inclusive collision. Among the most commonly used models are the Colour-Singlet Model (CSM)~\cite{Chang:1979nn,Berger:1980ni,Baier:1983va} and the Non-Relativistic QCD (NRQCD) factorisation~\cite{Bodwin:1994jh}, which takes into account higher $Q\overline{Q}$ Fock states ${}^{2S+1}L^{[c]}_J$, where $J$ is the total angular momentum, $S$ is the spin and $L$ is the orbital angular momentum of the quarkonium states. For the production of the $S$-wave quarkonia in NRQCD, the leading contribution of the series expansion on the relative $Q\overline{Q}$ velocity $v$ is the CSM ( ${}^3S^{[1]}_1$) and the next-to-leading-$v$ contributions ( ${}^1S^{[8]}_0$, ${}^3S^{[8]}_1$, ${}^3P^{[8]}_J$) result in what is referred to as the Colour-Octet Mechanism (COM).

The main motivation for the present work is to revisit the so-called "LEP II $J/\psi$ production puzzle": the inclusive NRQCD NLO calculation~\cite{PhysRevD.84.051501} for $J/\psi$ production in $\gamma\gamma$ fusion does not reproduce the experimental data from the DELPHI \cite{DELPHI:2003hen} experiment. On the experimental side, the dataset has only 16 events with $p_T>1$ GeV \cite{DELPHI:2003hen} and it was not confirmed by other experiments. We stress that the normalisation of the DELPHI data is unpublished~\cite{Todorova-Nova:2001hjh}, only the self-normalised distribution was published in \cite{DELPHI:2003hen}. Nevertheless, the LO calculation~\cite{Klasen:2001cu} for $J/\psi$ inclusive production in $\alpha_s$ NRQCD reproduced the DELPHI data. However, at NLO in $\alpha_s$, the NRQCD calculation \cite{PhysRevD.84.051501} is suppressed by a negative interference between the ${}^1S^{[8]}_0$ and ${}^3P^{[8]}_J$ contributions. As a result, the DELPHI data systematically overshoot the inclusive-$J/\psi$-production predictions at NLO in $\alpha_s$ in NRQCD. The CSM model exhibits a more perturbatively stable behaviour, which motivates our choice to revisit first this process in the CSM.

In this study, we take into account all sizeable direct and resolved-photon contributions. These are: 1) pure QED process $\gamma + \gamma \rightarrow J/\psi({}^3S^{[1]}_1) + \gamma$ up to $\alpha^3\alpha_s$, 2) single-resolved-photon contributions up to $\alpha \alpha_s^3$, 3) $\gamma+\gamma\to J/\psi + c\bar{c}$ up to $\alpha^2\alpha_s$ and 4) $\gamma+\gamma\to J/\psi + ggg$ up to $\alpha ^2 \alpha _s^3$.

The leading contribution of the direct-photon production, $\gamma+\gamma \rightarrow J/\psi + c\bar{c}$, has been computed up to NLO in $\alpha_s$ in the CSM in \cite{Chen:2016hju}. This dominant direct contribution is however not sufficient to reproduce the data, if we trust the normalisation of the data from~\cite{Todorova-Nova:2001hjh}, but it is a significant contribution to inclusive production in $\gamma \gamma$ fusion in the CSM. The NLO $K$-factor for this contribution has been found \cite{Chen:2016hju} to be close to unity, therefore in the present study we take this contribution at LO in $\alpha_s$. This and the other CS contributions have been put together for the first time in the present work, where we give phenomenological predictions for the kinematics of LEP II DELPHI and of the future CEPC experiments.

\vspace*{-0.7cm}
\section{Methodology of the cross-section computation}

\vspace*{-0.15cm}

\begin{figure}[h!]
    \centering
 \begin{subfigure}[b]{0.1\textwidth}
  \centering
  \includegraphics[width=1\textwidth]{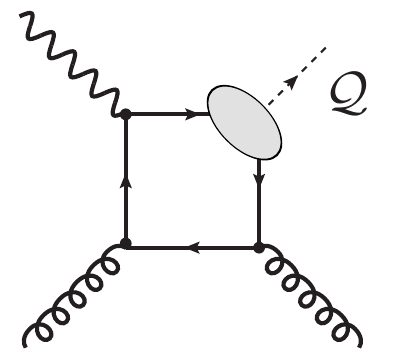}\vspace*{-0.25cm}
  \caption{   }
  \label{lo}
\end{subfigure} \hspace*{-0.35cm}
 \begin{subfigure}[b]{0.1\textwidth}
  \centering
  \includegraphics[width=1\textwidth]{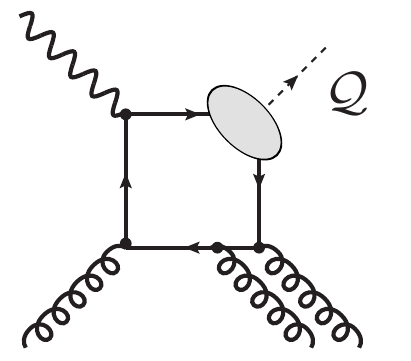}\vspace*{-0.25cm}
  \caption{   }
\label{q_gl}
\end{subfigure} \hspace*{-0.35cm}
 \begin{subfigure}[b]{0.1\textwidth}
  \centering
  \includegraphics[width=1\textwidth]{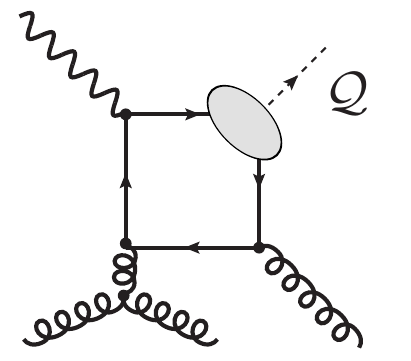}\vspace*{-0.25cm}
  \caption{   }
\label{init_gl_gl}
\end{subfigure} \hspace*{-0.35cm}
 \begin{subfigure}[b]{0.1\textwidth}
  \centering
  \includegraphics[width=1\textwidth]{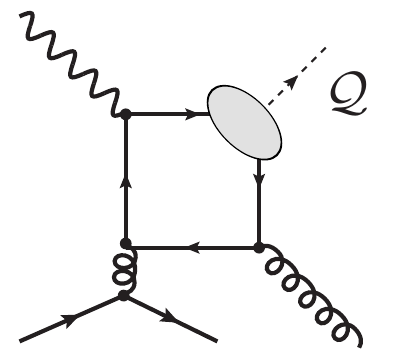}\vspace*{-0.25cm}
  \caption{   }
\label{init_gl_qq}
\end{subfigure} \hspace*{-0.35cm}
 \begin{subfigure}[b]{0.1\textwidth}
  \centering
  \includegraphics[width=1\textwidth]{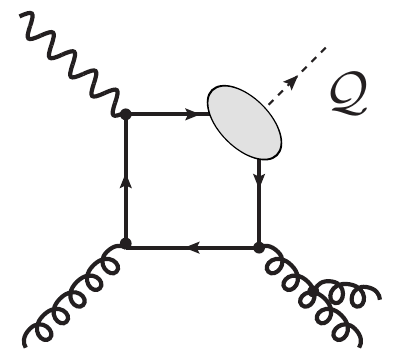}\vspace*{-0.25cm}
    \caption{   }
\label{fin_gl_gl}
\end{subfigure} \hspace*{-0.35cm}
 \begin{subfigure}[b]{0.1\textwidth}
  \centering
  \includegraphics[width=1\textwidth]{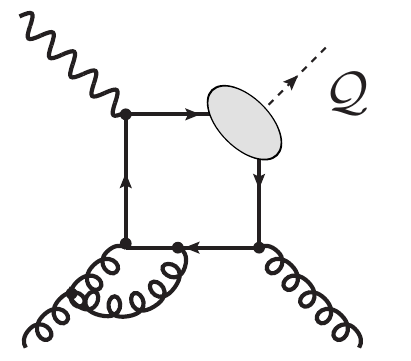}\vspace*{-0.25cm}
    \caption{   }
\label{vtx_corr}
\end{subfigure} \hspace*{-0.35cm}
 \begin{subfigure}[b]{0.1\textwidth}
  \centering
  \includegraphics[width=1\textwidth]{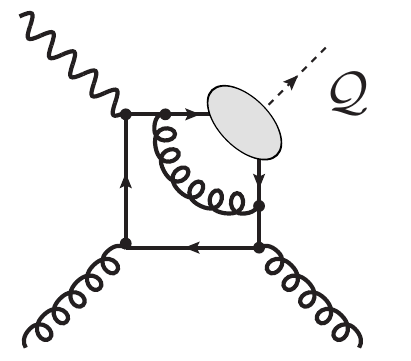}\vspace*{-0.25cm}
    \caption{   }
\label{box}
\end{subfigure} \hspace*{-0.35cm}
 \begin{subfigure}[b]{0.1\textwidth}
  \centering
  \includegraphics[width=1\textwidth]{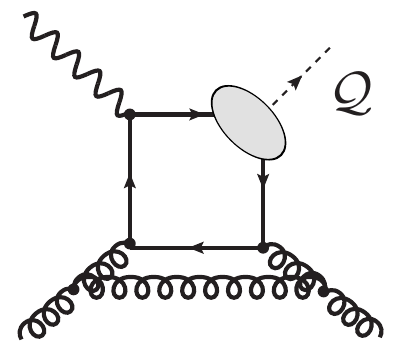}\vspace*{-0.25cm}
    \caption{   }
\label{loop}
\end{subfigure} \hspace*{-0.35cm}
 \begin{subfigure}[b]{0.1\textwidth}
  \centering
  \includegraphics[width=1\textwidth]{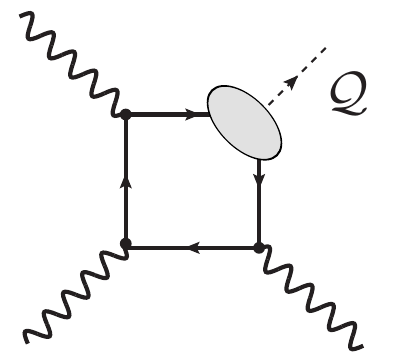}\vspace*{-0.25cm}
    \caption{   }
\label{box-ee}
\end{subfigure} \hspace*{-0.35cm}
 \begin{subfigure}[b]{0.1\textwidth}
  \centering
  \includegraphics[width=1\textwidth]{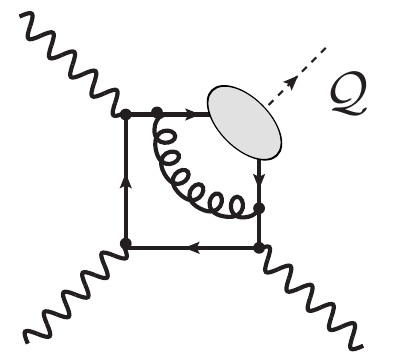}\vspace*{-0.25cm}
    \caption{   }
\label{loop-ee}
\end{subfigure} \hspace*{-0.35cm} \vspace*{-0.35cm}
\caption{An illustrative selection of Feynman diagrams for inclusive $\cal{Q}$ production in $\gamma\gamma$ fusion via CS channels for single-resolved-photon and direct-photon processes. The quark and anti-quark attached to the blobs are taken as on-shell and their relative velocity $v$ is set to zero.} 
\label{diagrams-CSM}
\end{figure}

\cf{diagrams-CSM} displays different types of Feynman diagrams for $\cal{Q}$ production in $\gamma \gamma$ fusion via CS channels for single-resolved-photon contributions at orders $\alpha \alpha_s^2$ (\cf{lo}), $\alpha \alpha_s^3$ (\cf{q_gl} -- \ref{loop}), and for direct-photon processes at orders $\alpha^3$ (\cf{box-ee}) and $\alpha^3\alpha_s$ (\cf{loop-ee}). To compute the cross section, one assumes two types of factorisation. The first one is the collinear one, by virtue of which the cross section can be written as the convolution of the photon flux in the Weizs\"acker Williams approximation~\cite{Frixione:1993yw} (and photon Parton Distribution Functions (PDFs) for the case of resolved-photon contributions) with the hard-scattering cross section $d\hat{\sigma}_{\gamma \gamma}$ (or $d\hat{\sigma}_{\gamma i}$ with $i=g,q$). The second factorisation concerns the hadronisation of the heavy-quark pair into the physical quarkonium state ${\cal Q}$. The partonic cross section is written as a product of the hard part that describes the $Q\overline{Q}$ pair production, and the soft part, via a non-perturbative matrix element, which describes the hadronisation of the $Q\overline{Q}$ pair into the physical quarkonium $\cal{Q}$. In the CSM, the latter one is proportional to the squared radial wave function at the origin $|R(0)|^2$ in the configuration space.

\vspace*{-0.2cm}
 
\vspace*{-0.15cm}

We have computed all these contributions using our codes based on {\it Wolfram Mathematica} packages \texttt{FeynArts}~\cite{Hahn:2000kx}, \texttt{FeynCalc}~\cite{MERTIG1991345} and \texttt{FormCalc}~\cite{Hahn:1998yk} for symbolic manipulation of the amplitudes. The \texttt{KIRA}~\cite{Maierhofer:2017gsa} package was employed to reduce the one-loop Feynman integrals to master integrals, which were numerically evaluated using the \texttt{LoopTools}~\cite{Hahn:1998yk} library. Finally, the \texttt{CUBA}~\cite{Hahn:2004fe} Monte-Carlo integration routines in \texttt{FORTRAN} were used to perform the final phase-space integration and the integration over the momentum fraction carried by the initial photon inside the corresponding lepton. The QCD Born-order multileg contributions 3) and 4) were computed with \texttt{HELAC-Onia}~\cite{Shao:2012iz}.

\vspace*{-0.3cm}
\section{Phenomenological study of quarkonium production in $\gamma \gamma$ fusion}
\vspace*{-0.15cm}

In \cf{figpt2Jpsi}, we present results of our computations of the above-mentioned direct and single-resolved-photon CS contributions for LEP II DELPHI (\cf{pt2-PRD79}) and the CEPC (\cf{pt2-CEPC}) kinematical conditions. The CS 1-loop QED direct-$\gamma$ contributions for $J/\psi$ have been evaluated for these kinematical conditions in the present contribution for the first time. The single-resolved-photon contributions are also considered for the first time for the CEPC kinematics. \cf{figpt2Jpsi} shows the $p_T^2$ dependence of the differential cross section integrated over rapidity $y\in [-2,2]$ in the positron-electron c.m.s. frame with the c.m.s. energy of the $e^+e^-$: $\sqrt{s_{ee}}=197$ GeV for LEP II DELPHI and $\sqrt{s_{ee}}=250$ GeV for the CEPC and the maximum scattering angle $\theta_{max}=25$~mrad. For the computation of the single-resolved-photon contributions, we have used the GRVGHO $\msbar$ photon PDF set~\cite{Gluck:1991jc}. In this analysis, we have taken $|R_{J/\psi}(0)|^2=$ {1.25}~GeV$^3$ and $m_c=1.5$~GeV, with the mass of quarkonium within NRQCD $M_{\cal{Q}}=2m_Q$. We have added a $20\%$ feed-down contribution from $\psi'$ decay (like in~\cite{Flore:2020jau}). The black data points in \cf{pt2-PRD79} are the DELPHI data. In \cf{figpt2Jpsi} and \cf{pt2-CEPC-Y}, the red dashed (solid) lines are the NLO (LO) single-resolved-photon predictions and the blue dashed (solid) lines are the corresponding NLO (LO) direct-photon predictions. The multileg CS channels $J/\psi + ggg$ and $J/\psi + c\bar{c}$ are plotted with green and violet lines correspondingly. The $J/\psi+ ggg$ contribution turns out to be small and this is why we do not show its uncertainty, while the QED and the single-resolved-photon contributions are relevant at low $p_T$, with respect to the $J/\psi + c\bar{c}$ contribution. The uncertainty bands about the corresponding curves in \cf{figpt2Jpsi} result from the $\mu_F$ and $\mu_R$ boundary scale values $\xi_{R,F}\in[0.5;2]$, where $\xi_{R,F}\equiv \mu_{R,F}/m_T$ and $m_T=\sqrt{M_{\cal{Q}}^2+p_T^2}$.

\vspace*{-0.15cm}
\begin{figure}[h!]
    \centering
 \begin{subfigure}[b]{0.47\textwidth}
  \centering
  \includegraphics[width=1\textwidth]{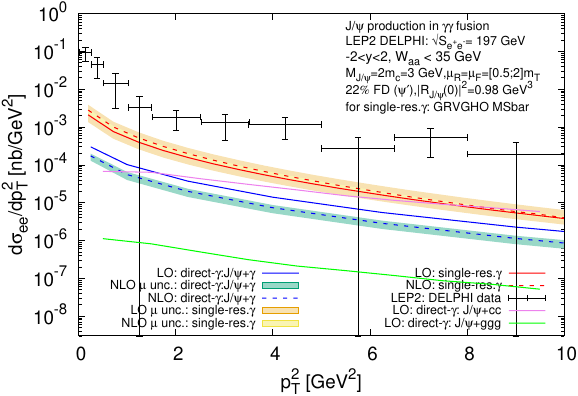}\vspace*{-0.2cm}
  \caption{   }
  \label{pt2-PRD79}
\end{subfigure}
\begin{subfigure}[b]{0.47\textwidth}
  \centering
  \includegraphics[width=1\textwidth]{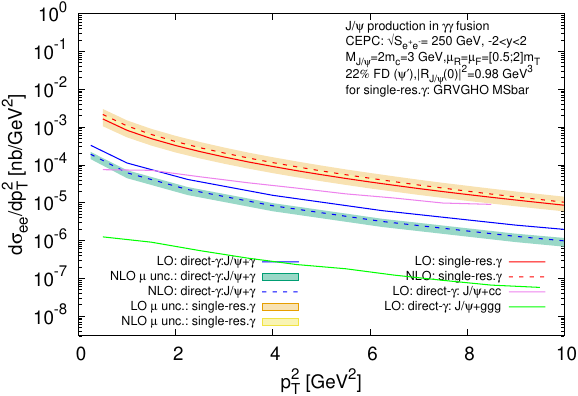}\vspace*{-0.2cm}
  \caption{   }
  \label{pt2-CEPC}
\end{subfigure}\vspace*{-0.3cm}
\caption{The $p_T^2$-differential cross section of inclusive $J/\psi$ production at LEP II DELPHI (\ref{pt2-PRD79}) and the CEPC (\ref{pt2-CEPC}). For the description of various curves see the main text.}
\label{figpt2Jpsi}
\end{figure}
\vspace*{-0.15cm}

In \cf{pt2-CEPC-Y}, we present the cross-section calculations for the CS 1-loop QED direct-photon and single-resolved-photon production of $\Upsilon(1S)$ mesons in the CEPC kinematical conditions which, to our knowledge, have been performed for the first time. For our predictions for $\Upsilon(1S)$ production, we have used $|R_{\Upsilon(1S)}(0)|^2=$ 7.5 GeV$^3$ and the $b$-quark mass $m_b=4.75$~GeV. We have estimated~\cite{ColpaniSerri:2021bla} the feed-down contributions from $\Upsilon(2S)$ to be 12.5$\%$ and from $\Upsilon (3S)$ to be 2.2$\%$. In \cf{pt2-CEPC-Y} we have adopted the same colour code for the different production channels as in \cf{figpt2Jpsi}. For bottomonium production, the CO contributions should be smaller than for charmonium production due to their $v$ scaling in NRQCD~\cite{Bodwin:1994jh}, which makes our predictions for $\Upsilon (1S)$ likely more complete in comparison with those for $J/\psi$.
\vspace*{-0.1cm}
\begin{figure}[h!]
\centering  
\includegraphics[width=0.47\textwidth]{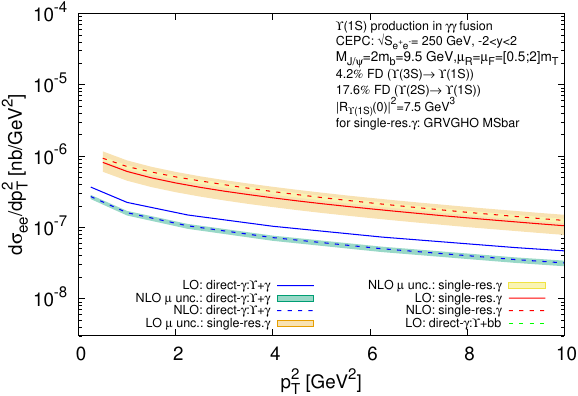}\\ \vspace*{-0.2cm}
\caption{The $p_T^2$-differential cross section of inclusive $\Upsilon$ production in $\gamma \gamma$ fusion in the CEPC kinematics. The notation for the different curves is the same as in \cf{figpt2Jpsi}.}
\label{pt2-CEPC-Y}
\end{figure}

\vspace*{-0.6cm}

Another outcome of our study is that the observation of exclusive $\gamma + \gamma \rightarrow J/\psi+\gamma$ process is within the LHC reach. We have estimated the $J/\psi$ production cross section in UPCs in the CMS detector for the following kinematic constraints: $\sqrt{s}=5.02$ TeV, $1.2<|y^{\psi}|<2.4$, $|\eta^{\gamma}|<2.4$, $p_T^{\psi}>2.5$ GeV, $p_T^{\gamma}>2.5$ GeV, to be about $O(10)$ nb. The photon-detection efficiency is expected to be close to unity if the trigger is associated with a $J/\psi$ so we can estimate the expected event counts to be $O(10)$ events with an expected integrated luminosity of ${\cal L}=13$ fb$^{-1}$. As a result, we can conclude that the exclusive $J/\psi + \gamma$ production can be measured in UPCs at the LHC. This gives us confidence that {\it inclusive} $J/\psi+\gamma$ and $J/\psi+X$ production from photon fusion, with likely much larger cross sections can be measured at the LHC too.

\vspace*{-0.35cm}
\section{Conclusions and outlook}
\vspace*{-0.15cm}

In this work, we have revisited the inclusive $e^+ e^- \rightarrow J/\psi + X$ process in the CSM. We have taken into account several direct-photon processes ($J/\psi + \gamma$, $J/\psi + c\bar{c}$) which we have found to be not negligible, especially in view of the fact that the normalisation of the DELPHI data has remained unpublished~\cite{Todorova-Nova:2001hjh}. We have presented the first predictions for the CSM one-loop QED direct-photon $J/\psi$ production for the CEPC and LEP II DELPHI kinematics and single-resolved-photon contributions for $J/\psi$ production at the CEPC. We also have provided the first predictions for the CSM one-loop QED direct-photon and single-resolved-photon contributions to $\Upsilon(1S)$ production at the CEPC. Additionally, our calculations indicate that exclusive $J/\psi + \gamma$ production can be measured in ultra-peripheral PbPb collisions at the LHC. 

\vspace*{-0.35cm}
\acknowledgments
\vspace*{-0.15cm}

We thank D. d'Enterria and K. Lynch for useful discussions. This project has received funding from the European Union’s Horizon 2020 research and innovation programme under grant agreement No. 824093 in order to contribute to the EU Virtual Access NLOAccess and a Marie Skłodowska-Curie action “RadCor4HEF” under grant agreement No. 101065263. This project has also received funding from the Agence Nationale de la Recherche (ANR) via the grant ANR-20-CE31-0015 (``PrecisOnium'') and via the IDEX Paris-Saclay "Investissements d’Avenir" (ANR-11-IDEX-0003-01) through the GLUODYNAMICS project funded by the "P2IO LabEx (ANR-10-LABX-0038)" and through the Joint PhD Programme of Universit\'e Paris-Saclay (ADI). YY acknowledges the financial support from the School of Physics, UCD. This work was supported by the French CNRS via the IN2P3 project GLUE@NLO and via the Franco-Polish EIA (GlueGraph).

\vspace*{-0.5cm}

\bibliographystyle{JHEP}
\bibliography{bibliography_021022}

\end{document}